\documentstyle[12pt,aaspp4]{article}
\def\eq#1{\begin{equation} #1 \end{equation}}

\def\about            {\hbox{$\sim$}}

\def\comm#1           {{\tt (COMMENT: #1)}}
\begin{document}

\title{ The Selection of RR Lyrae Stars Using Single-epoch Data  }

\author{
\v{Z}eljko Ivezi\'{c}\altaffilmark{\ref{Princeton},\ref{HNR}},
A. Katherina Vivas\altaffilmark{\ref{Venezuela}},
Robert H. Lupton\altaffilmark{\ref{Princeton}},
Robert Zinn\altaffilmark{\ref{Yale}}
}

\newcounter{address}
\setcounter{address}{1} \altaffiltext{\theaddress}{ Princeton University
Observatory, Princeton, NJ 08544 \label{Princeton}}
\addtocounter{address}{1}
\altaffiltext{\theaddress}{H.N. Russell Fellow, on leave from the University of Washington \label{HNR}}
\addtocounter{address}{1} \altaffiltext{\theaddress}{Centro de Investigaciones
de Astronom{\'\i}a (CIDA). Apartado Postal 264. M\'erida 5101-A. 
Venezuela \label{Venezuela}}
\addtocounter{address}{1}
\altaffiltext{\theaddress}{Yale University. Department of Astronomy.
PO BOX 208101. New Haven, CT 06511 \label{Yale}}

\begin{abstract}
We utilize a complete sample of RR Lyrae stars discovered by the QUEST
survey using light curves to design selection criteria
based on SDSS colors. Thanks to the sensitivity of the $u-g$ color
to surface gravity and of $g-r$ color to effective temperature, and
to the small photometric errors (\about 0.02 mag) delivered by SDSS,
RR Lyrae stars can be efficiently and robustly recognized even with
single-epoch data. In a 100\% complete color-selected sample,
the selection efficiency (the fraction of RR Lyrae stars in the
candidate sample) is 6\%, and, by adjusting color cuts, it can be 
increased to 10\% with a completeness of 80\%, and to 60\% with 28\% 
completeness. Such color selection produces samples that are 
sufficiently clean for statistical studies of the Milky Way's halo 
substructure, and we utilize it to select 3,643 candidate RR Lyrae
stars from SDSS Data Release 1. We demonstrate that this sample
recovers known clumps of RR Lyrae stars associated with the Sgr dwarf 
tidal tail, and Pal 5 globular cluster, and use it to constrain the
halo substructure away from the Sgr dwarf tidal tail. These results 
suggest that it will be possible to study the halo substructure 
out to $\sim$70 kpc from the Galactic Center in the entire area 
imaged by the SDSS, and not only in the multiply observed regions.
\end{abstract}


\section{Introduction}

Studies of substructures, such as clumps and streams, in the Galactic
halo can help constrain the formation history of the Milky Way. 
Hierarchical models of galaxy formation predict that these substructures 
should be ubiquitous in the outer halo, where the dynamical timescales 
are sufficiently long for them to remain spatially coherent (Johnston 
{\em et al.}~1996; Mayer {\em et al.}~2002, Helmi 2002). One of the best 
tracers to study the outer halo are RR Lyrae stars because

\begin{itemize}
\item
They are nearly standard candles (dispersion of $\sim0.13$ mag, Vivas et 
al. 2001) and thus it is straightforward to determine their distance, and
\item
They are sufficiently bright ($\langle M_V \rangle$ = 0.7-0.8,
\cite{L96}, Gould \& Popowski 1998) to be detected at large distances
(5-100 kpc for $14 < r < 20.7$).
\end{itemize}

RR Lyrae stars are typically found by obtaining well-sampled
light curves. The QUEST survey is the largest such survey that is
capable of discovering RR Lyrae stars in the outer halo. Using a
1m Schmidt telescope, the QUEST survey has so far discovered about
500 RR Lyrae stars in 400 deg$^2$ of sky (Vivas et al. 2003). 
Nevertheless, Ivezi\'{c} et al. (2000, hereafter I00) demonstrated that
RR Lyrae stars can be efficiently and robustly found even with two-epoch
data, using accurate multi-band photometry obtained by the Sloan
Digital Sky Survey (SDSS). The QUEST survey later demonstrated (Vivas 
et al. 2001) that most ($>$90\%) of the SDSS candidates are real RR Lyrae 
stars, and also confirmed the estimate of the sample completeness
($\sim35\pm5$\%).

To extend the above surveys for RR Lyrae stars to a significant
fraction of the sky (say, one quarter) is difficult. The QUEST survey
will cover up to 700 deg$^2$, while the SDSS survey, which should
observe close to one quarter of the sky, will obtain only 
single-epoch data for most of the scanned area. However, here we
demonstrate that the distinctive SDSS colors of RR Lyrae stars allow 
their selection using only a single-epoch of data. We utilize a complete 
sample of RR Lyrae stars discovered by the QUEST survey and design optimal 
selection criteria based on SDSS colors. The data and the selection method 
are described in Section 2, in Section 3 we select and analyze candidate
RR Lyrae stars from SDSS Data Release 1, and summarize and discuss the 
results in Section 4.

\section{The Selection of RR Lyrae Stars Using SDSS Colors }

\subsection{ The SDSS and QUEST Data }

The Sloan Digital Sky Survey (SDSS; York et al.~2000) is
revolutionizing studies of the Galactic halo because
it is providing homogeneous and deep ($r < 22.5$) photometry in five
passbands ($u$, $g$, $r$, $i$, and $z$, Fukugita et al.~1996;
Gunn et al.~1998; Smith et al.~2002; Hogg et al. 2002) accurate to
0.02 mag (Ivezi\'{c} et al.~2003a). The survey sky coverage of up to
10,000 deg$^2$ in the Northern Galactic Cap will result in
photometric measurements for over 100 million stars and a similar
number of galaxies. Astrometric positions are accurate to better than
0.1 arcsec per coordinate (rms) for sources with $r<20.5^m$
(Pier et al.~2003), and the morphological information from the images
allows reliable star-galaxy separation to $r \sim$ 21.5$^m$ (Lupton
et al.~2001).

Here we use SDSS imaging data which are part of the SDSS Data Release 1
(Abazajian et al.~2003, hereafter DR1). DR1 includes 2099 square 
degrees of five-band imaging data, to a depth of $r\sim22.6$.
SDSS equatorial observing runs 752 and 756 overlap with the 
QUEST observations in a 89 deg$^2$ large region defined by
$-1^\circ < \delta_{2000} < 0^\circ$, and
09$^h$ 44$^m$ $< \alpha_{2000} <$ 15$^h$ 40$^m$.
This region contains  about 210,000 unique, stationary unresolved 
sources with $14<r<20$, with mean galactic coordinates ($l=290^\circ$,
$b=53^\circ$). 
In the same region there are 162 RR Lyrae discovered
by the QUEST survey, and described by Vivas et al. (2001, 2003).
The RR Lyrae stars span a range of magnitudes of $V\sim 14-19.7$.
The discovery of these variables in this region of the sky
was based on high quality light curves, each containing  25 to 35 
different epochs. This sample of RR Lyrae stars has a high completeness 
($>90$\%) for the ab-type variables (fundamental mode pulsators). 
The completeness decreases for the low amplitude type c RR Lyrae stars
to $55-75$\%, depending on the magnitude of the star.

When computing the efficiency of the selection algorithms described below,
we exclude SDSS objects in the region 
$-0.58^\circ < \delta < -0.51^\circ$, which was not observed by QUEST 
because it fell on a gap between the columns of CCDs in the QUEST camera. 
All magnitudes have been corrected by interstellar extinction using 
the dust maps and transformations given by \cite{SFD98}.

\subsection{   The SDSS Observations of the QUEST RR Lyrae  }

We searched for the 162 QUEST RR Lyrae in the SDSS DR1 database\footnote{
Available from http://www.sdss.org} within
a circle of radius 2 arcsec centered on the QUEST position, and found 
all of them. The distribution of distances between the QUEST and SDSS 
positions has a median of 0.5 arcsec, and root-mean-square scatter of 
0.16 arcsec (the distributions of $\alpha_{2000}$ and $\delta_{2000}$
differences show offsets of 0.3 arcsec for each coordinate).

The SDSS processing flags (for details see DR1 and Stoughton et al. 2002)
indicated that 9 stars may have substandard photometry (complex blends, 
cosmic rays, bad pixels), with probable errors sometimes as large as 
0.05 mag. Since the selection algorithm discussed here relies
on accurate color measurements, hereafter we consider the sample of
153 stars with impeccable photometry.

\begin{figure}
\plotfiddle{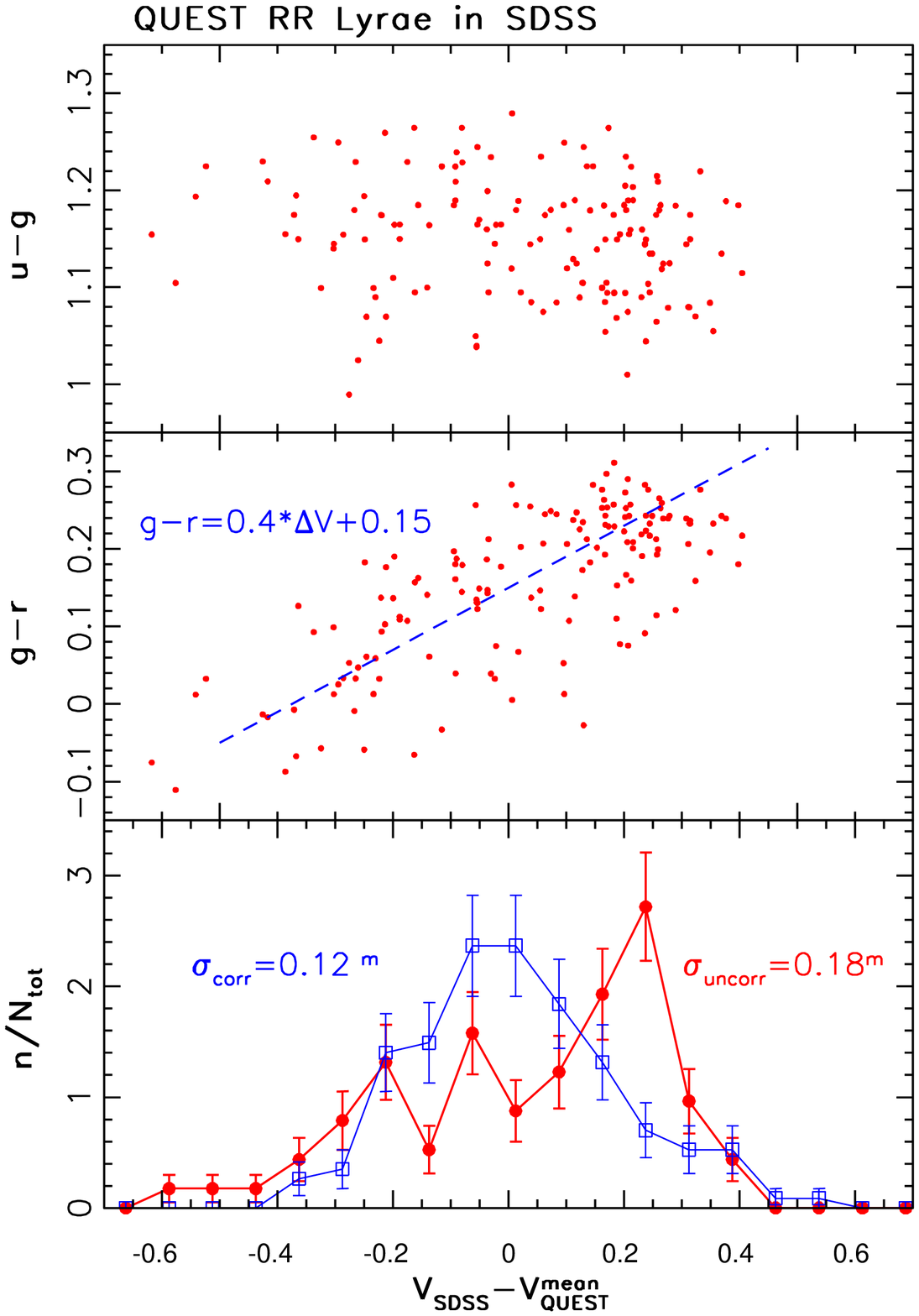}{12cm}{0}{65}{65}{-200}{-50}
\caption{The top and middle panels show the correlations between 
the single-epoch $u-g$ and $g-r$ colors measured by SDSS, and the 
difference between the mean $V$ magnitude measured by the QUEST 
survey ($V_{QUEST}^{mean}$) and a single-epoch synthetic $V$ 
magnitude measured by the SDSS ($V_{SDSS}$), for 153 RR Lyrae 
stars observed by both surveys. Note that RR Lyrae stars have 
bluer $g-r$ colors when brighter, while there  is no discernible 
correlation for $u-g$ color. The dashed line in the middle panel 
shows a best-fit relation between the $g-r$ color and 
$V_{SDSS}-V_{QUEST}^{mean}$. The bottom panel compares the 
distribution of $V_{SDSS}-V_{QUEST}^{mean}$ differences (solid
circles) to the distribution of differences when $V_{SDSS}$ is
corrected for this correlation (open squares, see eq.~\ref{Vrrl}).
\label{deltaV}}
\end{figure}

The histogram marked by solid circles in the bottom panel in Figure 
\ref{deltaV} shows the distribution of differences between the mean $V$ 
magnitude measured by the QUEST survey, $V_{QUEST}^{mean}$, and a 
single-epoch synthetic SDSS-based $V_{SDSS}$ magnitude, computed from
(Fukugita et al. 1996) 
\eq{
             V_{SDSS}  - r = 0.44\,(g-r) - 0.02.
}
Reassuringly, the mean value of the shown distribution is consistent with
zero to within 0.02 mag. The distribution is skewed because RR Lyrae stars
have asymmetric light curves (they spend more than 50\% of their variability
cycle fainter than their mean magnitude).

The top and middle panels show the correlations between the $u-g$ and
$g-r$ colors measured by SDSS and the $V$ magnitude difference. As expected,
RR Lyrae stars have bluer $g-r$ colors when brighter, while there is no discernible
correlation for $u-g$ color, which may be due to shock wave related activity 
(Smith 1995). Note that the $g-r$ color spans twice as large a range as does 
the $u-g$ color.

The $g-r$ color is correlated with $V_{SDSS}-V_{QUEST}^{mean}$. The best-fit
relation, shown in the middle panel by the dashed line is
\eq{
             g-r = 0.4\,(V_{SDSS}-V_{QUEST}^{mean}) + 0.15.
}
This relation can be used to correct a bias in single-epoch SDSS measurements 
due to unknown phase, such that 
\eq{
\label{Vrrl}
             V_{SDSS}^{RRLyr} = r - 2.06\,(g-r) + 0.355,
}
where all measurements have been corrected for ISM reddening.  
The histogram marked by open squares in the bottom panel in Figure 
\ref{deltaV} shows the distribution of $V_{SDSS}^{RRLyr}-V_{QUEST}^{mean}$.
The root-mean-square scatter is significantly decreased compared to the
scatter in $V_{SDSS}-V_{QUEST}^{mean}$ (0.12 mag. vs. 0.18 mag., as marked 
in the figure). This is a relation that produces unbiased RR Lyrae distances
with a minimal scatter (0.12 mag.) from single-epoch SDSS measurements. 
It is remarkable that the scatter in mean magnitude estimated from single-epoch 
SDSS measurements is as small as the intrinsic uncertainty in RR Lyrae absolute 
magnitudes. We note that assuming a constant $M_r$ (instead of $M_V$) to 
determine distances results in practically no bias, but the scatter is 
increased to 0.20 mag. (for constant $M_g$ the bias is 0.23 mag. with a 
comparable scatter).

\subsection{The Colors of RR Lyrae in SDSS Photometric System}

\begin{figure}
\plotfiddle{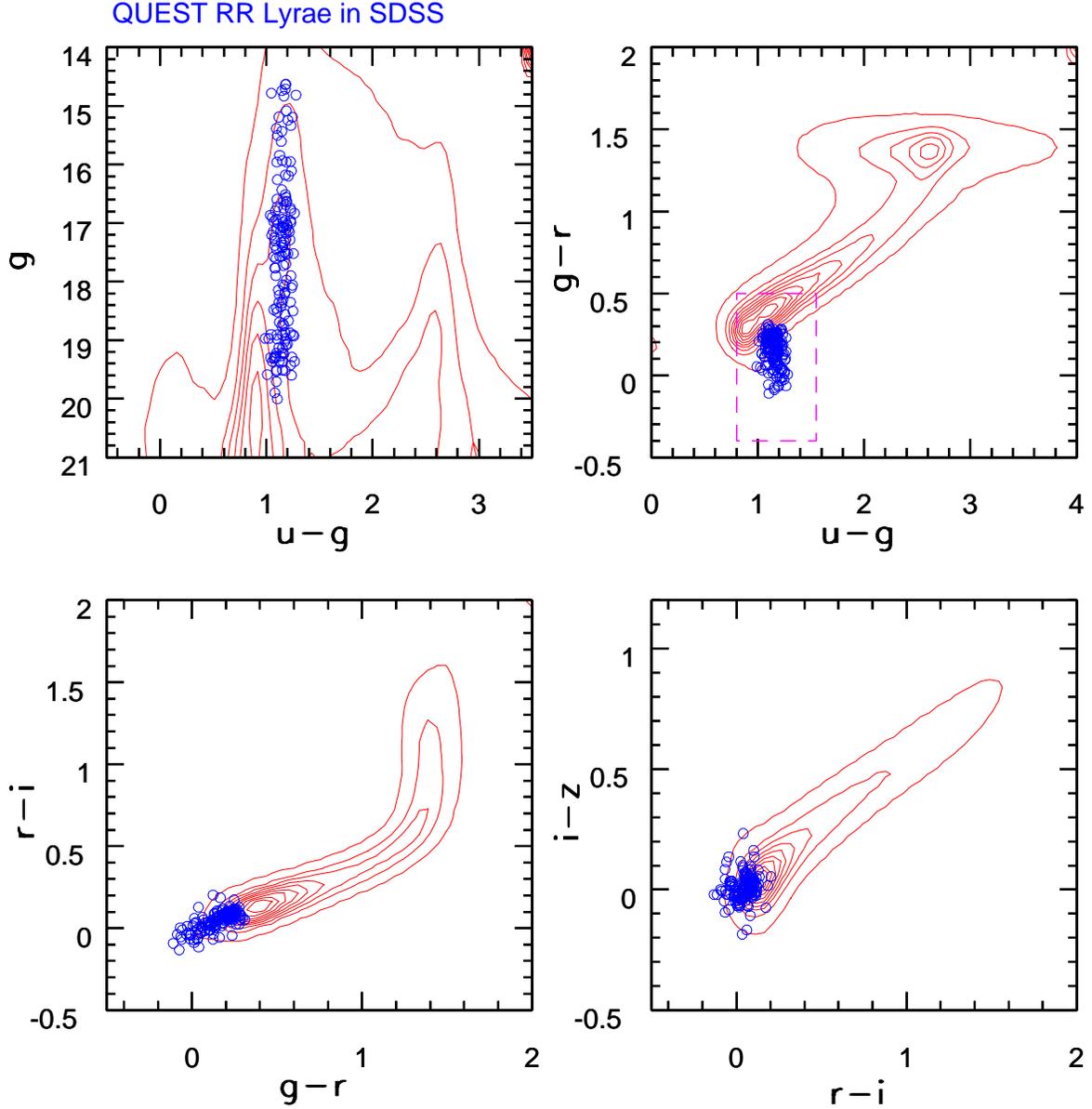}{13cm}{0}{85}{85}{-250}{-50}
\caption{The comparison of the distribution of point sources in
the SDSS color-magnitude and color-color diagrams (linearly spaced
contours) and the distribution of RR Lyrae stars (symbols). The
symbol size corresponds to 3-5 times the photometric errors,
depending on the scale of individual panels. The rectangle shown
by the dashed lines in the top right panel is the region which
is shown magnified in Figure \ref{RRLyraeBoxes}. Note that RR Lyrae
stars span a very narrow range of $u-g$ color ($u-g \sim 0.3\pm0.06$).
\label{compareCCD}}
\end{figure}

Figure \ref{compareCCD} shows the distribution of all point sources
with $r<20$ in the SDSS color-magnitude and color-color diagrams as 
linearly spaced contours. In color-color diagrams, red is always towards 
the upper right. For a detailed description of stellar colors in the SDSS
photometric system see Finlator et al. (2000) and references
therein. The 153 QUEST RR Lyrae stars are shown as symbols. The
symbol size corresponds to 3-5 times the photometric errors,
depending on the scale of individual panels (for a detailed analysis of 
SDSS photometric errors see Ivezi\'{c} et al. 2003a). That is, the scatter
of points is due to intrinsic differences among RR Lyrae stars and
the variation of colors with phase.
Nevertheless, RR Lyrae stars span a very narrow range of SDSS
colors. The color limits for the sample discussed here are

\begin{eqnarray}
          0.99 < u-g < 1.28 \\
         -0.11 < g-r < 0.31 \\
         -0.13 < r-i < 0.20 \\
         -0.19 < i-z < 0.23
\end{eqnarray}

In particular, both the range ($\about 0.30$ mag) and the root-mean-square 
scatter ($\about 0.06$ mag) are the smallest for $u-g$ color.

The RR Lyrae fraction is 1 in 1,300 amongst all SDSS stars with $r<20$.
In a subsample selected using the above color cuts, the RR Lyrae fraction
is 6\%. We show next how this fraction can be increased to over 60\%
by optimizing the color selection boundaries.

\subsection{The Optimization of the Color Selection}

\begin{figure}
\plotfiddle{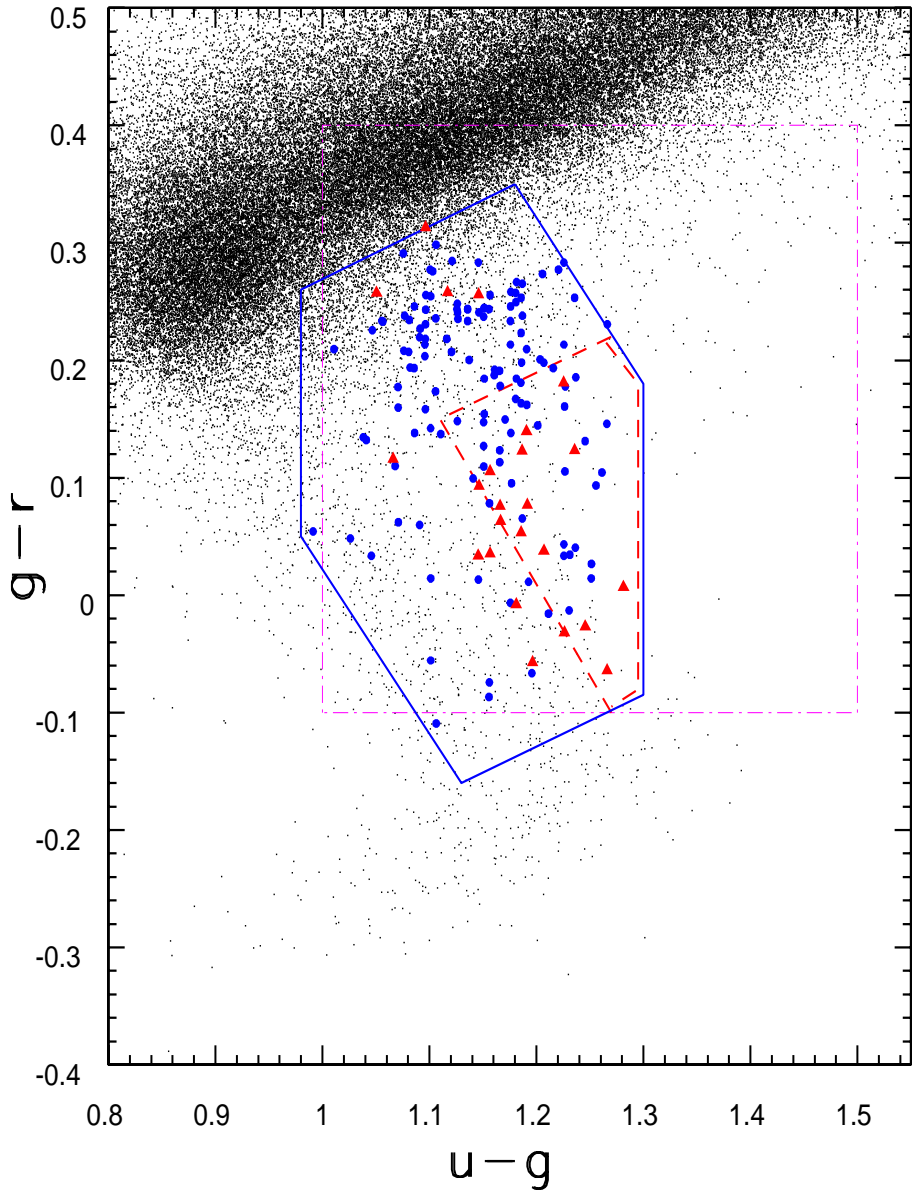}{14.0cm}{0}{75}{85}{-250}{-70}
\caption{The selection criteria for RR Lyrae stars. The small dots
show all SDSS point sources with $r<20$, and the large symbols are
confirmed RR Lyrae stars (solid circles are stars of the type $ab$ and
triangles are $c$ type). The photometric errors are comparable
to the radius of the large dots. The solid polygon is a suggested boundary
for the 100\% completeness, with efficiency of 6\%. The dashed lines
are an example of a restricted selection boundary which results
in a completeness of 28\% and 61\% efficiency. The dot-dashed lines
are the selection boundary from a variability study by Ivezi\'{c} (2000), 
shown here for reference.
\label{RRLyraeBoxes}}
\end{figure}

RR Lyrae stars are found furthest from the locus of other stars in the
$g-r$ vs. $u-g$ color-color diagram. The relevant part of this
diagram, outlined by the small rectangle in Figure \ref{compareCCD},
is shown magnified in Figure \ref{RRLyraeBoxes}. The small dots
show all SDSS point sources with $14<r<20$, and the symbols are
confirmed RR Lyrae stars (solid circles are $ab$ type RR Lyrae stars and triangles
correspond to the $c$ type stars). The photometric errors are comparable
to the radius of the large circles, except for objects in the
top left corner which have faint $u$ band magnitudes ($\la 21.5$).

Most of the contamination in a sample of candidate RR Lyrae stars
selected using simple color cuts listed above comes from the main
stellar locus visible in the top left corner. The second most
significant source of contamination is the A star locus running
from the main locus towards the bottom right. This motivates a revised
selection boundaries shown as the polygon outlined by solid lines.
The edges with positive $d(g-r)/d(u-g)$ slope have a constant
distance from the main stellar locus
\eq{
       D_{ug} = (u-g) + 0.67\,(g-r) - 1.07,
}
and the edges with negative slope have a constant
distance from the A star locus
\eq{
         D_{gr} = 0.45\,(u-g) - (g-r) - 0.12.
}
The QUEST RR Lyrae span the ranges $-0.05 < D_{ug} < 0.35$ and
$0.06 < D_{gr} < 0.55$, resulting in a selection efficiency
(the fraction of RR Lyrae stars among the selected candidates)
of 6\%.

The selection efficiency can be further increased by increasing
the lower limits on $D_{ug}$ and $D_{gr}$, $D_{ug}^{Min}$ and
$D_{gr}^{Min}$, respectively (of course, the sample completeness
then becomes less than 100\%), and keeping the restrictions
in colors $(u-g)$, $(r-i)$ and $(i-z)$. For example, the dashed lines
show a restricted selection boundary obtained with
$D_{ug}^{Min}=0.15$ and $D_{gr}^{Min}=0.23$, which results
in a completeness of 28\% and 61\% efficiency\footnote{The
efficiency quoted here is obtained with $r<20$ and no limit on
$u$. Since the faintest QUEST RR Lyrae star in the sample has
$u=21.2$ and $r=19.7$, the selection efficiency could be increased
by a factor of 1.2 by requiring $u < 21.1$ and $r < 19.7$.}.

\begin{figure}
\plotfiddle{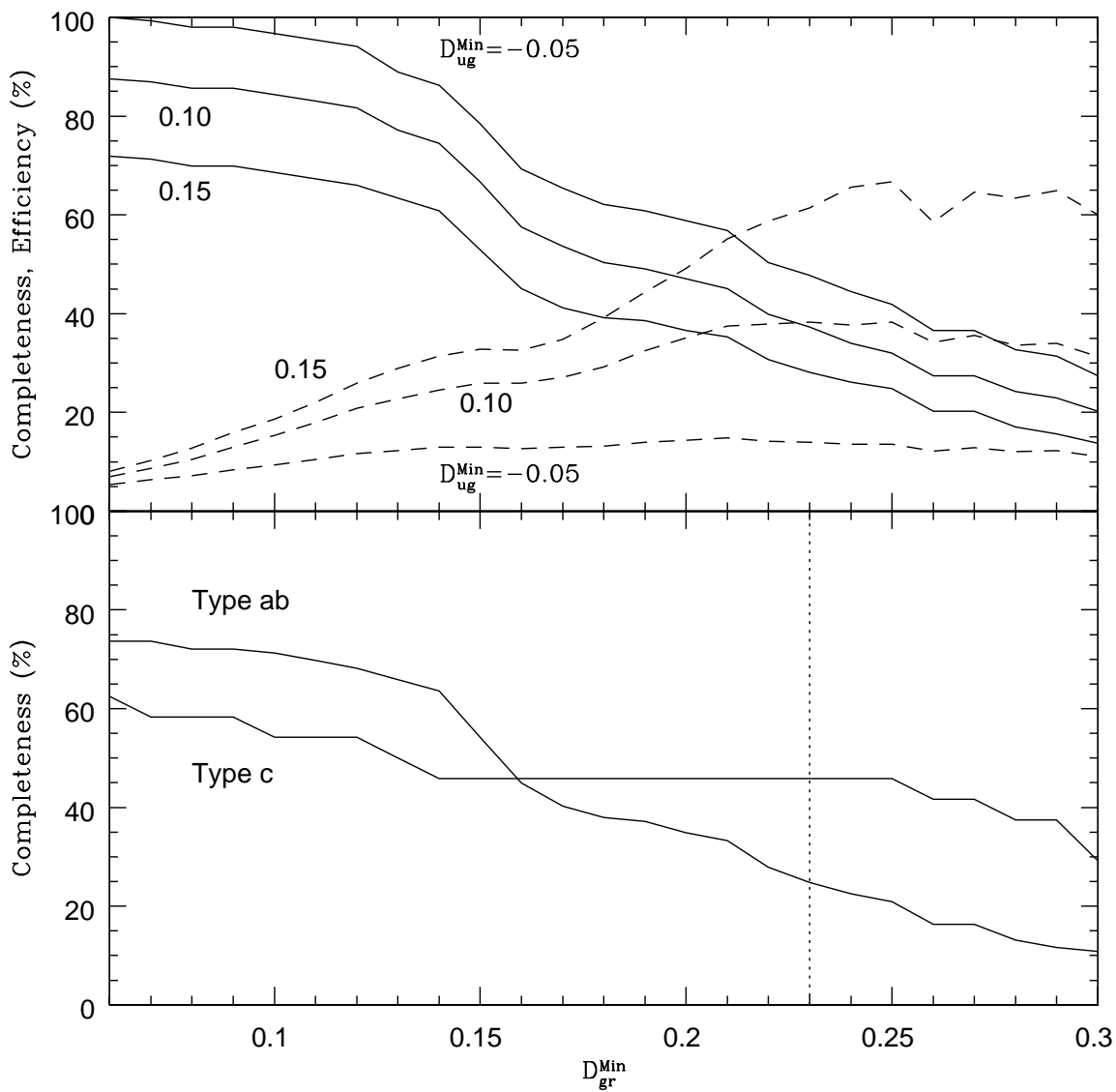}{13cm}{0}{80}{80}{-250}{-130}
\caption{The panel shows the dependence of the selection completeness 
(solid lines) and efficiency (dashed lines) as a function of $D_{ug}^{Min}$
(different curves, as labeled) and $D_{gr}^{Min}$.  The bottom panel
compares the completeness estimates for the different types of RR Lyrae 
variables for a color cut with $D_{ug}^{Min}=0.15$. The dotted line
marks a cut with $D_{gr}^{Min}=0.23$. Type $c$ RR Lyrae stars have a 
higher selection efficiency for $D_{gr}^{Min}> 0.16$ than type $ab$ 
RR Lyrae.
\label{ComplEff}}
\end{figure}

The top panel in Figure \ref{ComplEff} shows a detailed dependence of the
completeness (solid lines) and efficiency (dashed lines) as
functions of $D_{gr}^{Min}$ for three different values of
$D_{ug}^{Min}$, as indicated. The largest possible selection
efficiency using single-epoch data is about 65\%. The remaining
contaminants are probably dominated by quasars (Richards et al.
2001) and non-variable horizontal branch stars.

The restricted color criteria preferentially select type $c$ RR Lyrae 
stars because they have bluer $g-r$ colors than $ab$ type stars (see
Figure~\ref{RRLyraeBoxes}).
The bottom panel in Figure \ref{ComplEff} compares the completeness 
estimates for the two different types of RR Lyrae variables 
for a color cut with $D_{ug}^{Min}=0.15$. When $D_{gr}^{Min}> 0.16$,
type $c$ RR Lyrae stars have a higher selection efficiency. For example,
in the region enclosed by the dashed lines in Figure \ref{RRLyraeBoxes}, 
we  recover 45\% of all QUEST type $c$ RR Lyrae stars but only 
25\% of the more common types $ab$.

To summarize, the proposed selection criteria are
\begin{eqnarray}
             14 < r < 20  \\
          0.98 < u-g < 1.30    \\
        D_{ug}^{Min}  < D_{ug} < 0.35 \\
        D_{gr}^{Min}  < D_{gr} < 0.55 \\
          -0.15 < r-i < 0.22 \\
          -0.21 < i-z < 0.25,
\end{eqnarray}
where $D_{ug}^{Min}$ and $D_{gr}^{Min}$ can be chosen using Figure
\ref{ComplEff} to yield desired selection completeness and
efficiency, depending on a specific purpose.

\section{    Candidate RR Lyrae Stars in SDSS Data Release 1    }

We apply the color selection method discussed in the preceeding Section 
to SDSS Data Release 1. DR1 includes sky regions with known halo 
substructures traced by RR Lyrae, and can thus be used to test 
the performance of the method for discovering such structures.
DR1 also includes areas for which variability data does not exist
(either from QUEST, or from repeated SDSS scans), and which may
exhibit previously uncharted substructure.

\begin{figure}
\plotfiddle{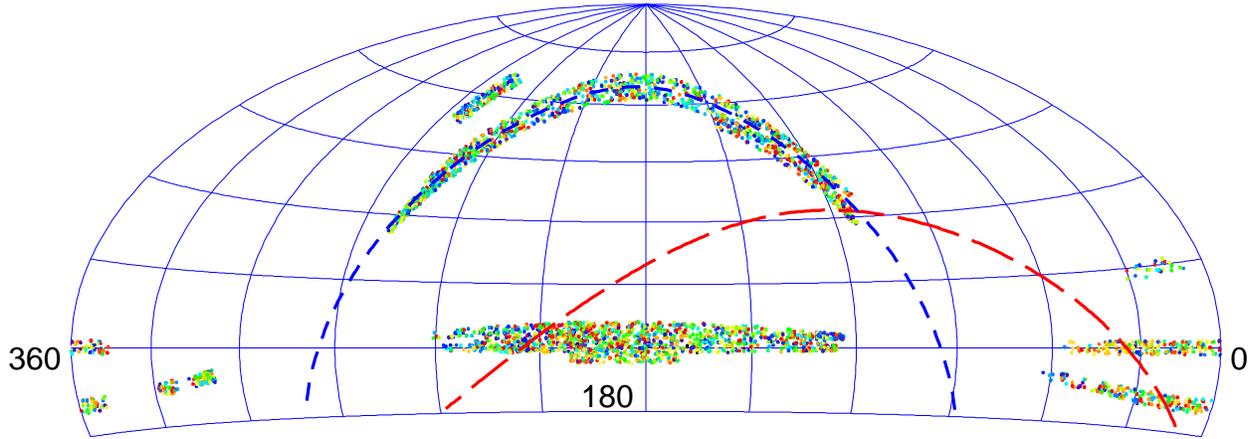}{3cm}{0}{90}{90}{-280}{-420}
\caption{The distribution of color-selected RR Lyrae 
candidates from SDSS Data Release 1, shown in Aitoff equatorial
projection. The long-dashed line indicates the position of the 
Sgr dwarf tidal stream, and the short-dashed line is a great circle 
that tracks a large fraction of the SDSS DR1 region (node=95$^\circ$,
inclination=65$^\circ$). The $r$ vs. RA 
distribution of stars along the latter great circle is shown in 
Fig.\ref{rvsRAgc}.
\label{Aitoff}}
\end{figure}

The condition  $D_{ug}^{Min}=0.10$ and $D_{gr}^{Min}=0.20$ results
in a sample of 3,643 candidate RR Lyrae stars selected from SDSS DR1
database. Here we require that the processing flags SATURATED and 
BRIGHT are not set, and use un-dereddened magnitudes for the
$14 < r < 20$ condition (of course, the color selection must be 
done with dereddened magnitudes). A subsample satisfying $D_{ug}^{Min}=0.15$ 
and $D_{gr}^{Min}=0.23$ contains 896 stars. The completeness and efficiency,
determined using Figure \ref{ComplEff}, for the first selection criteria 
are $C=50\%, E=35\%$, and for the second selection criteria 
$C=28\%, E=60\%$. The distribution of selected candidates on the 
sky is shown in Figure~\ref{Aitoff} (which closely outlines the SDSS DR1 
area).

\subsection{            A Self-consistency Test            }

The mean density of RR Lyrae stars can be estimated
from 
\begin{equation}
            \rho_{RRLyr} = {N_s \, E_s \over A_{DR1} C_s},
\end{equation}  
where the area included in DR1 is $A_{DR1}=2099$ deg$^2$, and $N_s$
is the number of selected stars using particular values of 
$D_{ug}^{Min}$ and $D_{gr}^{Min}$. The estimate $\rho_{RRLyr}$ should
be nearly the same for both samples, and should agree with the
value of $\rho_{RRLyr}\sim 1.3$ deg$^2$, determined from the QUEST
data for their first 400 deg$^2$ of sky (Vivas et al. 2003). 
The values obtained here, 1.21 deg$^2$ and 0.91 deg$^2$, are in 
good agreement with each other, indicating that the selection
method is robust (the training sample included data for a 24 times 
smaller area). 

We estimate that SDSS DR1 contains $\sim2200$ RR Lyrae stars, 
and that 1,170 are included in our sample of 3,643 candidates. 
The smaller, more restrictive, sample of 896 stars contains 540 
probable RR Lyrae stars.

\subsection{ A Test of the Ability to Recover Halo Substructure }

We analyze the spatial structure of selected candidates by examining
their distribution in the $r$ vs. position diagrams, for narrow
strips on the sky. The equatorial strip (Dec$\sim0^\circ$) contains
several known clumps of RR Lyrae stars (I00, Vivas et al. 2001). 

\begin{figure}
\plotfiddle{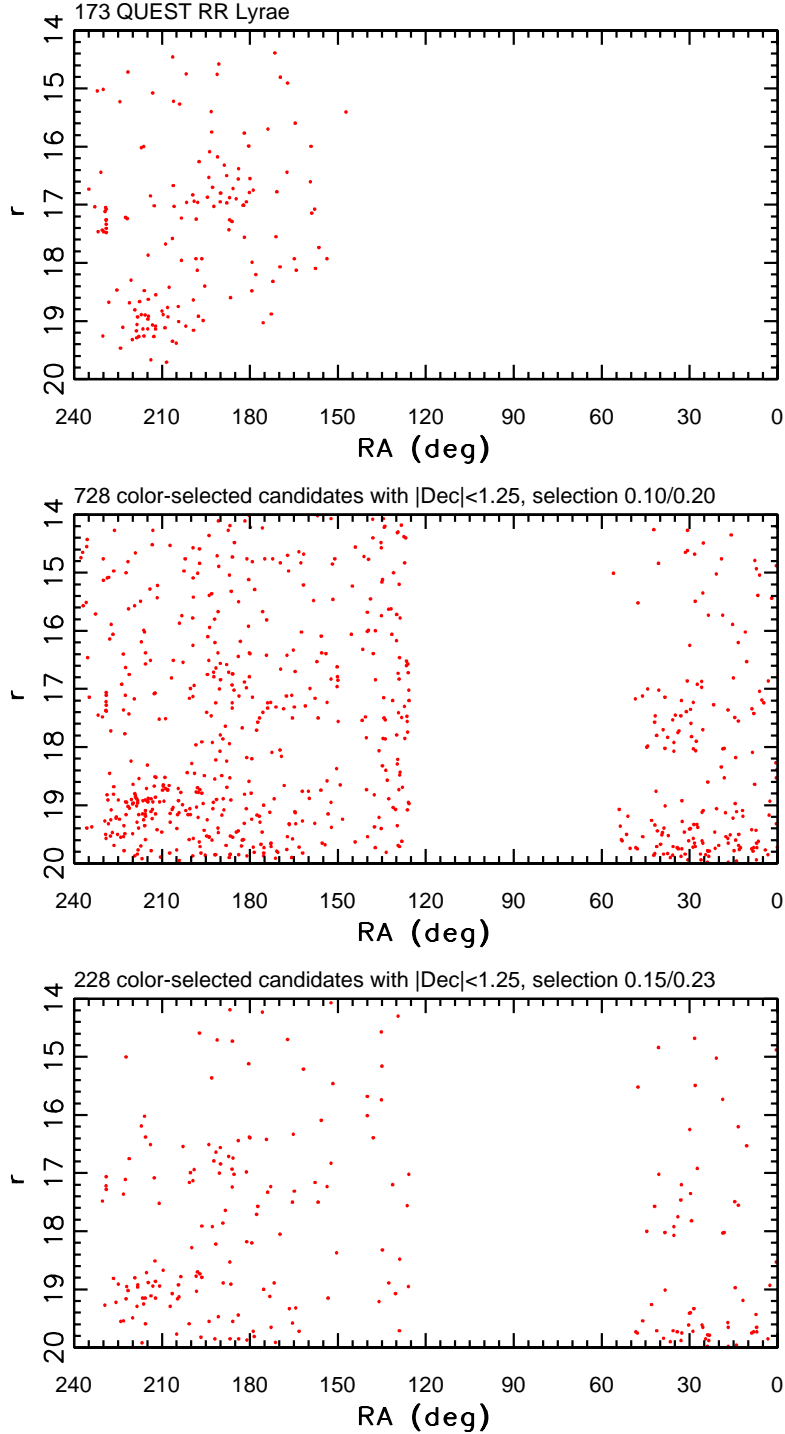}{18cm}{0}{110}{110}{-200}{-470}
\caption{The comparison of the $r$ vs. RA distribution of
QUEST RR Lyrae stars (top panel) and color-selected candidates
using SDSS single-epoch measurements (middle and bottom panels). 
The QUEST sample of confirmed RR Lyrae stars is practically complete
(in the region $150<$RA$<240$, while the estimated completeness and 
efficiency for SDSS samples are 50\%/35\% and 28\%/60\%, for the 
middle and bottom panels, respectively (in the sampled RA range). 
Note that the clumps associated with 
the Sgr dwarf tidal tail (RA \about 215, $r$ \about 19) and 
Pal 5 globular cluster (RA \about 230, $r$ \about 17.4), as well
as a clump at (RA \about 190, $r$ \about 17), are recovered by
color-selected SDSS samples. The clump at (RA \about 35, $r$ \about 
17.5) is also associated with the Sgr dwarf tidal tail.
\label{rvsRA}}
\end{figure}

The distribution of QUEST RR Lyrae stars in the $r$ vs. RA diagram
along the Celestial equator is shown in the top panel in Figure~\ref{rvsRA}.
The $r$ range from 14 to 20 corresponds to distances of 5 kpc to 70 kpc 
(recall that the strip width in the Dec direction for the QUEST subsample
is 1 deg). Three especially prominent features are the clump associated
with the Sgr dwarf tidal tail (RA \about 215, $r$ \about 19), Pal 5 globular 
cluster and associated tidal debris (RA \about 230, $r$ \about 17.4),
and a clump at (RA \about 190, $r$ \about 17).

All these three features are recovered by the color-selected SDSS DR1 
candidate RR Lyrae stars, whose $r$ vs. RA distributions are shown in
in the middle and bottom panels in Figure~\ref{rvsRA}. In particular,  the 
feature at (RA \about 190, $r$ \about 17), detected at a $5\sigma$ level
above the background by Vivas \& Zinn (2002, 2003) using a complete 
sample of confirmed RR Lyrae stars is clearly visible.

The color-selected samples also recover the so-called ``southern'' clump 
at RA \about 30 and $r$ \about 17--18,
(associated with the Sgr dwarf tidal tail, see the great circle marked by 
long-dashed line in Figure~\ref{Aitoff}) that was discovered using A-colored 
stars by Yanny et al. (2000). Furthermore, the faint clump at RA \about 30 
and $r > 19$ is present in a sample of candidate RR Lyrae stars selected
from repeated SDSS observations (Ivezi\'{c} et al. 2003b). 
Given these successful recoveries of known
structure, we conclude that the color-selected samples of candidate RR Lyrae
stars are sufficiently clean and robust to study halo substructure out to 
distances of $\sim$70 kpc.

\subsection{  Is Sgr Tidal Stream the Most Prominent Halo Feature? }

To reliably answer the question posed in the title of this subsection, 
one would need an all-sky survey of several halo tracers. 
While such data do not exist yet (the upcoming large scale synoptic 
surveys will eventually discover all halo RR Lyrae), a study of 2MASS
data by Majewski et al. (2003) provided the first all-sky view of 
halo structure, traced by M giants. They did not find any features
that would compete with the prominence of the  Sgr tidal stream.
Nevertheless, M giants are not as good standard candles as RR Lyrae
stars, and are more sensitive to metallicity effects. It is therefore
worthwhile to examine the distribution of candidate RR Lyrae stars in 
areas of sky that were not explored until now. 

\begin{figure}
\plotfiddle{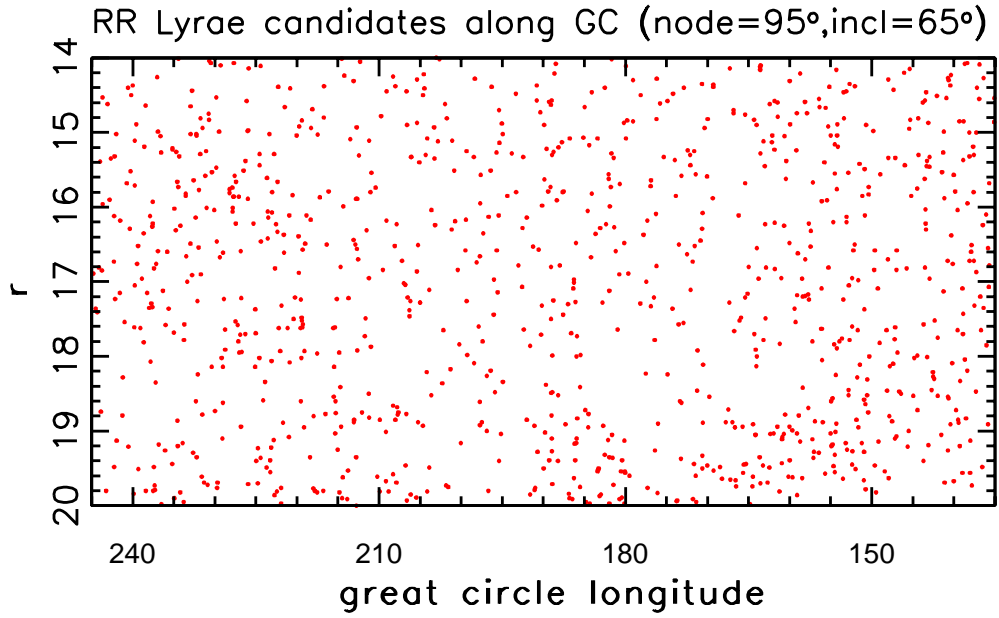}{3cm}{0}{80}{80}{-260}{-300}
\caption{The $r$ vs. great circle longitude distribution of SDSS
color-selected candidates (selection 0.10/0.20), along a great circle marked 
by the short-dashed line in Fig.~\ref{Aitoff}. The structure is not
as pronounced along this great circle, as it is along the Celestial
Equator (see the middle panel of Fig.~\ref{rvsRA}, note different
scale for x axis).
\label{rvsRAgc}}
\end{figure}

In Figure~\ref{rvsRAgc} we examine magnitude-angle diagrams for candidates 
selected using $D_{ug}^{Min}=0.10$ and $D_{gr}^{Min}=0.20$, along the 
great circle (within $\pm5^\circ$) marked by short-dashed line in Figure~\ref{Aitoff}, 
and defined by node=95$^\circ$ and inclination=65$^\circ$, relative to the Celestial 
Equator (for more details about great circle coordinates see Pier et al. 2003). 
The structure seen in this Figure is not nearly as prominent as that shown in the
middle panel of Figure~\ref{rvsRA}, where the candidate RR Lyrae stars were
selected by the same criteria. Two possible overdensities are visible at
the longitude of \about 225 and $r$ \about 16, and the longitude of \about 
150--165 and $r$ \about 19-20. The latter is supported by the distribution
of candidate RR Lyrae stars selected from repeated SDSS observations (and is
probably associated with the Sgr tidal stream, Ivezi\'{c} et al. 2003c), 
while such data are not available in the region of the former overdensity. 
We are currently investigating the distribution of candidate variable stars 
selected by comparing POSS and SDSS measurements (Sesar et al. 2003, in prep.) 
in order to derive more reliable conclusions about the halo substructure in that 
region.

\section{                        Discussion                 }

The robust recovery of the known halo substructures with the color selection proposed 
here suggests that it will be possible to constrain the halo structure out to $\sim70$ 
kpc from the Galactic Center in the entire area imaged by the SDSS, and not only in the 
multiply observed regions. This method may result in discoveries of more Sgr dwarf 
debris in currently unexplored parts of sky, which would be important to understand 
the evolution of the disruption of this galaxy. If events similar to the accretion and 
disruption of Sgr dwarf have occurred with other galaxies, this technique has a good chance 
of discovering their signatures. Therefore, before large-scale variability surveys,
such as Pan-STARRS and LSST become available, the candidate RR Lyrae stars selected using 
SDSS colors can be used for statistical studies of the halo substructure. The selection 
efficiency of \about60\%, with a completeness of \about30\%, should be sufficient to 
uncover the most prominent features.

\acknowledgments Observations for the QUEST RR Lyrae Survey were obtained at
the Llano del Hato National Observatory, which is operated by CIDA
for the Ministerio de Ciencia y Tecnolog{\'\i}a of Venezuela. 
Funding for the creation and distribution of the SDSS
Archive has been provided by the Alfred P. Sloan Foundation, the Participating
Institutions, the National Aeronautics and Space Administration, the
National Science Foundation, the U.S. Department of Energy, the
Japanese Monbukagakusho, and the Max Planck Society. The SDSS Web site
is http://www.sdss.org/.  The SDSS is managed by the Astrophysical
Research Consortium (ARC) for the Participating Institutions.  The
Participating Institutions are The University of Chicago, Fermilab,
the Institute for Advanced Study, the Japan Participation Group, The
Johns Hopkins University, Los Alamos National Laboratory, the
Max-Planck-Institute for Astronomy (MPIA), the Max-Planck-Institute
for Astrophysics (MPA), New Mexico State University, Princeton
University, the United States Naval Observatory, and the University of
Washington. AKV and RZ were partially funded by the National Science Foundation
under grant AST-0098428. \v{Z}I thanks Princeton University for generous 
financial support of this research.

%%%%%%%%%%%%%%%%%%%%%
%%%  BIBLIOGRAPHY %%%
%%%%%%%%%%%%%%%%%%%%%

\newpage
\clearpage
%%%%%%%%%%%%%%%%%%%%%
%%% FIGURES %%%%%%%%%
%%%%%%%%%%%%%%%%%%%%%

%%%%% Fig 1 %%%%%%

%%%%% Fig 2 %%%%%%

%%%%%% Fig 3 %%%%%%

%%%% Fig 4 %%%%%%

%%%%% Fig 5 %%%%%%

%%%%% Fig 6 %%%%%%

%%%%% Fig 7 %%%%%%


\begin{thebibliography}{}
\bibitem[]{} Abazajian, K., {\em et al.} 2003, AJ, 126, 2081
\bibitem[Finlator 2000]{F2000} Finlator, K., {\em et al.} 2000, AJ, 120, 2615 
\bibitem[Fukugita {\em et al.} 1996]{F96}Fukugita, M., Ichikawa, T., Gunn, J.E.,
         Doi, M., Shimasaku, K., \& Schneider, D.P. 1996, AJ, 111, 1748
\bibitem[]{} Gould, A., \& Popowski, P. 1998, ApJ, 508, 844
\bibitem[Gunn {\em et al.} 1998]{Gunnetal} Gunn, J.E., {\em et al.} 1998,
         AJ, 116, 3040
\bibitem[Helmi 2002]{} Helmi, A. 2002, Ap\&SS, 281, 351
\bibitem[]{hogg} Hogg, D.W., Finkbeiner, D.P., Schlegel, D.J. \& Gunn, J.E. 2002, AJ, 122, 2129
\bibitem[Ivezi\'c {\em et al.} 2000]{I00} Ivezi\'c, \v Z., {\em et al.}
         2000, AJ, 120, 963 (I00)
\bibitem[Ivezi\'c {\em et al.} 2003a]{I03a} Ivezi\'c, \v Z., {\em et al.}
         2003a, Proceedings of the Workshop {\it Variability with Wide Field Imagers}, 
         MemSAIt, in press (astro-ph/0301400)
\bibitem[Ivezi\'c {\em et al.} 2003b]{I03b} Ivezi\'c, \v Z., {\em et al.}
         2003b, Proceedings of the Conference ``Satellites and Tidal Streams'', May 26-30, 2003,
                      La Palma, Spain (also astro-ph/0309075)
\bibitem[Ivezi\'c {\em et al.} 2003c]{I03c} Ivezi\'c, \v Z., {\em et al.}
         2003c, Proceedings of the Conference ``Milky Way Surveys: The Structure and Evolution 
         of Our Galaxy'', June 15-17, 2003, Boston (also astro-ph/0309074)
\bibitem[Johnston et al. 1996]{kvj96} Johnston, K.V., Hernquist, L., \&
         Bolte, M., 1996, ApJ, 465, 278
\bibitem[Layden {\em et al.} 1996]{L96} Layden, A.C., Hanson, R.B., Hawley,
                          S.L., Klemola, A.R., \& Hanley, C.J. 1996, AJ, 112, 2110
\bibitem[]{} Lupton, R.H. {\it et al.} 2001, in {\it Astronomical Data Analysis Software and
           Systems X}, ASP Conference Proceedings, Vol.238, p. 269.
\bibitem[]{} Majewski, S., {\it et al.} 2003, astro-ph/0304198
\bibitem[Mayer et al. 2002]{mayer02} Mayer, L., Moore, B., Quinn, T.,
          Governato, F., \& Stadel, J., 2002, astro-ph/0110386
\bibitem[Pier {\em et al.} 2003]{Pier03} Pier, J.R. {\em et al.} 2003, AJ, 125, 1559
\bibitem[]{} Richards, G.T., {\em et al.} 2001, AJ, 121, 2308
\bibitem[Saha 1984]{S84} Saha, A. 1984, ApJ 283, 580
\bibitem[Schlegel, Finkbeiner \& Davis 1998]{SFD98} Schlegel, D.,
         Finkbeiner,D.P. \& Davis, M. 1998, ApJ 500, 525
\bibitem[]{smith} Smith, J.A. {\it et al.} 2002, AJ, 123, 2121
\bibitem[Stoughton {\em et al.} 2002]{EDR} Stoughton, C., {\em et al.}
         2002, AJ, 123, 485 (EDR)
\bibitem[]{} Vivas, A.K. et al. 2001, ApJL, 554, L33
\bibitem[]{} Vivas, A. K. \& Zinn R. 2002, Proceedings of the Workshop
{\it Variability with Wide Field Imagers}, MemSAIt, in press (astro-ph/0212116)
\bibitem[]{} Vivas, A.K. et al. 2003, AJ, submitted
\bibitem[]{} Vivas. A. K. \& Zinn, R. 2003, in preparation
\bibitem[]{} Yanny, B., {\em et al.} 2000, ApJ, 540, 825
\bibitem[York  {\em et al.} 2000]{York} York, D.G., {\em et al.} 2000,
         AJ, 120, 1579
\end{thebibliography}
\end{document}